\documentclass[12pt]{article}
\usepackage{amsfonts,amssymb}

\newtheorem{theorem}{Theorem}
\newtheorem{lemma}{Lemma}
\newtheorem{remark}{Remark}
\newtheorem{proposition}{Proposition}
\newtheorem{corollary}{Corollary}
\newcommand{\ju}[4]{\mbox{$
  \left(\begin{array}{cc}{#1} & {#2}\\{#3} &{#4}  \end{array}\right)$}}

\newcommand{\R}{\mathbb{R}}

\newcommand{\M}{\mathbb{M}}

\newcommand{\Z}{\mathbb{Z}}

\newcommand{\la}{\lambda}

\newcommand{\pa}{\partial}

\newcommand{\be}{\begin{equation}}
\newcommand{\ee}{\end{equation}}
\newcommand{\beq}{\begin{eqnarray}}
\newcommand{\eeq}{\end{eqnarray}}
\newcommand{\beqq}{\begin{eqnarray*}}
\newcommand{\eeqq}{\end{eqnarray*}}

\newcommand{\La}{\Lambda}

\title{\bf Negative order MKdV hierarchy and a new
 integrable Neumann-like system} 

\author{Zhijun Qiao$^{1,2,*}$\\
$^1$T-7 and CNLS, MS B-284, Los Alamos National Laboratory\\
Los Alamos, NM 87545, USA\\
$^2$Institute of Mathematics, Fudan University\\
 Shanghai 200433, PR
China\\
$^{*}${\footnotesize E-mail: qiao@cnls.lanl.gov \ \
zjqiao@yahoo.com}} \date{}

\begin{document}
\maketitle
\begin{abstract}
The purpose of this paper is to develop the negative order MKdV
hierarchy and to present a new related integrable Neumann-like
Hamiltonian flow 
from the view point of inverse recursion operator and constraint
method.  The whole MKdV hierarchy both positive and negative
is generated by the kernel elements of Lenard's operators pair and
recursion operator.  
Through solving a key operator equation, the whole 
MKdV hierarchy is shown to have the Lax representation. In particular, 
some new integrable equation together with the Liouville equations,
 the sine-Gordon equation, and the sinh-Gordon equation
are derived from the negative order MKdV hierarchy.  It is very
interesting that the restricted flow, corresponding to the negative
order MKdV hierarchy, is just a new kind of Neumann-like
system.  This new Neumann-like system is obtained through restricting the
MKdV spectral problem onto a symplectic submanifold and is proven to
be completely integrable under the Dirac-Poisson bracket, which we
define on the symplectic submanifold.  Finally, with the help of the
constraint between the Neumann-like system and the negative order MKdV
hierarchy, all equations in the hierarchy are proven to have the
parametric representations of solutions. In particular, we obtain the
parametric solutions of the sine-Gordon equation and the sinh-Gordon
equation.
\end{abstract}

   $ $\\
{\it AMS Subject: 35Q53; 58F07; 35Q35}

\section{Introduction}
\setcounter{equation}{0} It is well-known that the nonlinear evolution
equations (NLEEs) solved by the famous inverse scattering
transformation (IST) can be understood as compatible conditions of
some linear equations \cite{ZMNP,AS}, namely, Lax representation.  In
the past two decades the Lax representation has played a very
important role in the discussion of NLEEs. In particular, the Lax
representation has been used successively in the bi-Hamiltonian
structure of finite-dimensional dynamical systems \cite{BN,AFW}, in
the nonlinearization theory of soliton system to produce new
completely integrable systems in the Liouville sense \cite{C1,C2}, in
the tri-Hamiltonian formulation of nonlinear equations \cite{FH}, and
in the finite-dimensional restricted flows of the underlying infinite
systems \cite{AW}. Recently, the finding of peaked solitons produced a
breakthrough in the study of nonlinear partial differential equations
\cite{CH}, where Camassa and Holm showed their equation is completely
integrable by the IST method. This fact allows one to discuss the
nonlinear dynamics of soliton solutions and billiard solution
\cite{ACFHM} via their linear spectral content (i.e. Lax
representation).  Thus, for a given hierarchy of NLEEs, to find the
Lax representation is of great importance.
    
The modified Korteweg-de Vries (MKdV) equation together with the MKdV
hierarchy has wide applications in physics and other nonlinear
sciences. It possesses the Lax pair \cite{AS}, the periodic soliton
solution \cite{Wa,Wa1}, the bi-Hamiltonian structure \cite{CP} and
other soliton properties such as Darboux transformation, B\"acklund
transformation and the Miura transformation between it and the KdV
equation \cite{O}.  About the study of the MKdV hierarchy, i.e. usual
higher-order MKdV equations, there have been many discussions in the
literature. In Ref. \cite{CR}, the sine-Gordon equation and the
intergrated MKdV equations were shown to conserve the same infinite set
of charges which were determined by a recursion relation. Afterward,
Verosky \cite{VJM} introduced the negative powers of Olver recursion operator
and presented the relations between the sine-Gordon/sinh-Gordon
equation and the potential MKdV equations (non-local flows). In 1993, 
Andree and Shmakova \cite{AS1} discussed
the supersymmetry structure of the sine-Gordon equation,  which can be
embraced in the MKdV and KdV hierarchies. All of those interesting
facts happened between the sine-Gordon equation and the MKdV equation.
The fact that the sine-Gordon equation is a negative flow
of MKdV can be seen in Ref. \cite{ZMNP}.

In recent years, the time-discrete version \cite{CN} of integrable
systems have already arisen a lot of attractive attentions.  This idea
was applied to constructing a new Lax pair from the old Lax pair
\cite{Qph}. In Ref. \cite{Qam}, we proposed an approach to generate
the positive and the negative order integrable hierarchies from a
given spectral problem. In fact, we had this idea in Ref.  
\cite{Qcaj}. But that depends on the existence of inverse recursion
operator. This means, for every concrete spectral problem, we need to
determine the inverse of recursion operator.

In the present paper, we will give the inverse of recursion operator of the
MKdV hierarchy in an explicit form. With the help of the
recursion operator and its inverse, we present the positive order and 
the negative order MKdV hierarchies of
NLEEs. The whole MKdV hierarchy is proven to have the Lax pair
through employing a key operator equation, and is
therefore a completely integrable hierarchy.
Particularly, some new integrable equation together with the Liouville equations,
 the sine-Gordon equation, and the sinh-Gordon equation
are derived from the negative order MKdV hierarchy.  Furthermore, the constraint between the MKdV
spectral problem and the negative order MKdV hierarchy derives a
restricted Hamiltonian flow on a symplectic submanifold, which is a
new kind of Neumann-like system.  Then, we define the Dirac-Poisson
bracket on the symplectic submanifold, which is a very useful tool to
deal with the finite dimensional integrable system on some
submanifols. Under the Dirac-Poisson bracket, the new Neumann-like system
has the Hamiltonian canonical form and has furthermore independent and
involutive functions, which guarantees its complete integrability.
Finally, each equation in the negative order MKdV hierarchy is proven
to have the parametric representation of solution, which is given by
the involutive solution of Hamiltonian phase flows (i.e. $x$-flow and
$t_m$-flow, $m<0, m\in \Z$).  Particularly, we
obtain the parametric solutions of the sine-Gordon equation and the
sinh-Gordon equation.

The whole paper is organized as follows.  Next section gives the pair
of Lenard's operators, the recursion operator and their inverses, and
introduces a key operator equation which is available for both the
positive order and the negative order MKdV hierarchies. This operator
equation is different from the one usually considered in literatures
\cite{C0,QDAKNS,Qlevi}.  In section 3, we will see that the positive
order MKdV hierarchy, arising from the kernel of one of Lenard's
operators, is nothing but the well-known MKdV hierarchy. In section 4,
we present the negative order MKdV hierarchy by using the kernel
element of the other one of Lenard's operators. All equations in the
hierarchy have the Lax pairs. In sections 5 and 6, we provide a new
kind of integrable Neumann-like system on a symplectic submanifold, and
give the parametric solutions of the negative order MKdV hierarchy,
respectively. In the last section, we give some conclusions.

For convenience, we make the following conventions:
\begin{eqnarray*}
f^{(m)}=\left\{\begin{array}{ll}
               \frac{\pa^m}{\pa x^m}f=f_{mx}, & m=0,1,2,...,\\
               \underbrace{\int\ldots\int}_{-m}f dx, & m=-1,-2,...,
               \end{array}
         \right.
\end{eqnarray*}
$f_t=\frac{\pa f}{\pa t}$, $f_{mxt}=\frac{\pa^{m+1} f}{\pa t\pa x^m}\
(m=0,1,2,...)$, $\pa=\frac{\pa}{\pa x}$, $\pa^{-1}$ is the inverse of
$\pa$, i.e.  $\pa \pa^{-1}=\pa^{-1}\pa=1$, $\pa^mf$ means the operator
$\pa^mf$ acts on some function $g$, i.e.
\begin{eqnarray*}
\pa^mf\cdot g=\pa^m(fg)
=\left\{\begin{array}{ll}
               \frac{\pa^m}{\pa x^m}(fg)=(fg)_{mx}, & m=0,1,2...,\\
               \underbrace{\int\ldots\int}_{-m}fg dx, & m=-1,-2,....
               \end{array}
         \right.
\end{eqnarray*}

In the following the function $u$ stands for potential, the imaginary
unit $i$ is satisfying $i^2=-1$, and $\la$ is assumed to be a spectral
parameter, and the domain of the spatial variable $x$ is $\Omega$
which becomes equal to $(-\infty,\ +\infty)$ or $(0, \ T)$, while the
domain of the time variable $t_m$ is the positive time axis
$\R^{+}=\left\{t_m\right|\ t_m\in \R, \ t_m\geq 0, \
m=0,\pm1,\pm2,...\}$.  In the case $\Omega=(-\infty,\ +\infty )$, the
decaying condition at infinity and in the case $\Omega=(0,\ T)$, the
periodicity condition for the potential function is imposed.

    $(\R^{2N},dp\wedge dq)$ stands for the standard symplectic
structure in Euclid space $\R^{2N}=\left\{\left.(p,q)\right| \
p=(p_1,\ldots,p_N), q=(q_1,\ldots,q_N)\right\}$, $p_j,q_j$
$(j=1,\ldots,N)$ are $N$ pairs of canonical coordinates,
$\left<\cdot,\cdot\right>$ is the standard inner product in $\R^N$; in
$(\R^{2N},dp\wedge dq)$, the Poisson bracket of two Hamiltonian
functions $F,H$ is defined by \cite{AV}
\begin{equation}
 \left\{F,H\right\}=\sum_{j=1}^N\left(\frac{\partial F}{\partial
q_j}\frac{\partial H}{\partial p_j} -\frac{\partial F}{\partial
p_j}\frac{\partial H}{\partial q_j}\right)=\left<\frac{\partial
F}{\partial q},\frac{\partial H}{\partial p}\right>
-\left<\frac{\partial F}{\partial p},\frac{\partial H}{\partial
q}\right>. \label{PB}
\end{equation}
$\la_1,...,\la_N$ are $N$ distinct spectral parameters, and
$\La=diag(\lambda _1,...,\lambda _N)$.  Denote all infinitely times
differentiable functions on real field $\R$ and all integers by
$C^{\infty}(\R)$ and by $\Z$, respectively.

\section{Inverse recursion operator and
operator equation}

\setcounter{equation}{0}

Let us consider the following spectral problem \be
\psi_x=\left(\begin{array}{lr} -i \la & u\\ u & i \la
                       \end{array}\right)\psi,  \ \ \ \ \psi=
\left(\begin{array}{l} \psi_1\\ \psi_2
                       \end{array}\right),  \label{6.4.0}
\ee which is quite a special case of the well-known
Zakharov-Shabat-AKNS spectral problem \cite{ZS} \be
\psi_x=\ju{-i\la}{u}{v}{i\la}\psi \ee with $v=u$.

Eq. (\ref{6.4.0}) is equivalent to \be
L\cdot\psi\equiv\ju{i\pa}{-iu}{iu}{-i\pa}\cdot\psi =\la\psi
\label{ZS0}\ee and its spectral gradients
$\nabla\la\equiv\frac{\delta\la}{\delta u} =\psi^2_2-\psi^2_1$
satisfies the Lenard eigenvalue problem \be K\cdot\nabla\la=\la^2
J\cdot\nabla\la
  \label{6.4.2}
\ee with the Lenard's operators pair \beq K=-\frac{1}{4}\pa^3+\pa
u\pa^{-1}u \pa, \ \ \ J =\pa,
         \label{6.4.3}
\eeq which yields the recursion operator \be {\cal
L}=J^{-1}K=-\frac{1}{4}\pa^2+ u\pa^{-1}u \pa.
              \label{6.4.4} \ee

Apparently, the Gateaux derivative operator $L_{*}(\xi)$ of the
spectral operator $L$ given by Eq. (\ref{ZS0}) in the direction
$\xi\in C^{\infty}(\R)$ is \be
L_{*}(\xi)\stackrel{\triangle}{=}\left.\frac{d}{d \epsilon}
\right|_{\epsilon=0}L(u+\epsilon\xi) =\left(\begin{array}{lr} 0 & -i
\xi\\ i\xi & 0
                       \end{array}\right)
                        \label{6.4.5}
\ee which is obviously an injective homomorphism.

Through guesswork and calculations, we can obtain the inverse
operators of $L,\ J,\ K$ and ${\cal L}$: \beq L^{-1} & = &
\ju{A}{-\pa^{-1}uA}{\pa^{-1}uA}{-A}, \label{6.4.6}\\ & & A=
-ie^{u^{(-1)}}\pa^{-1} e^{-2u^{(-1)}}u\pa^{-1}e^{u^{(-1)}}\pa u^{-1},
\nonumber\\ J^{-1} & = & \pa^{-1}, \label{6.4.7}\\ K^{-1} & = &
-4\pa^{-1}e^{-2u^{(-1)}}\pa^{-1}
e^{4u^{(-1)}}u\pa^{-1}e^{-2u^{(-1)}}\pa u^{-1}\pa^{-1},
             \label{6.4.8}\\
  {\cal L}^{-1} & = & -4\pa^{-1}e^{-2u^{(-1)}}\pa^{-1}
               e^{4u^{(-1)}}u\pa^{-1}e^{-2u^{(-1)}}\pa u^{-1}.
                         \label{6.4.9}
\eeq

   For any given $C^{\infty}$-function $G$, we construct the following
    operator equation with respect to $V=V(G)$ \beq
    [V,L]&=&L_{*}(K\cdot G)-L_{*}(J\cdot G)L^2. \label{VLMKDV} \eeq

\begin{remark}  This equation contains a special term $L^2$ instead of the
  term $L$ usually considered in literatures \cite{C0,QDAKNS,Qlevi}.
  \end{remark}

  \begin{theorem}
For the MKdV spectral operator (\ref{ZS0}) and an arbitrarily given
$C^{\infty}$-function $G$, the operator equation (\ref{VLMKDV}) has
the following solution \beq
V&=&V(G)=\ju{(uG_x)^{(-1)}\pa-\frac{1}{2}uG_x}
{\frac{1}{2}G_x\pa-\frac{1}{4}G_{xx}}
{\frac{1}{2}G_x\pa-\frac{1}{4}G_{xx}}
{(uG_x)^{(-1)}\pa-\frac{1}{2}uG_x}. \label{6.4.15} \eeq
\label{Th1}
\end{theorem}
{\bf Proof}: \ \ Directly substituting Eqs. (\ref{6.4.15}),
(\ref{ZS0}), (\ref{6.4.3}) and (\ref{6.4.5}) into Eq. (\ref{VLMKDV}),
we can complete the proof of this theorem.

$ $
\section{Positive order hierarchy of Eq. (\ref{6.4.0}), i.e.
         usual MKdV hierarchy} \setcounter{equation}{0} Let us now
give the positive order MKdV hierarchy through considering the kernel
element of the Lenard's operator $J$.

 $G_0=a\in Ker J$ and the recursion operator (\ref{6.4.4}) yield
the positive order hierarchy of Eq. (\ref{ZS0}) 

\be u_{t_m} =J{\cal
L}^m\cdot a, \ \ m=0,1,2,\ldots, \label{6.4.19} \ee 
which has the
following representative equations \beq u_{t_1}=au_x, \ \ {\rm trivial
\ case}, \label{6.4.20} \\
u_{t_2}=-\frac{1}{4}au_{xxx}+\frac{3}{2}au^2u_x. \label{6.4.21} \eeq
Here, $a=a(t_n)\in C^{\infty}(\R)$ is an arbitrarily
given fuction with respect to variables $t_n \ (n\geq 0, n\in \Z)$, but
independent of $x$. 
Apparently, with $a=4$ Eq. (\ref{6.4.21}) becomes the well-known MKdV
equation \be u_{t_2}-6u^2u_x+u_{xxx}=0. \label{6.4.21k} \ee Therefore,
Eq. (\ref{6.4.19}) coincides with the well-known MKdV hierarchy, which
corresponds to the isospectral case: $\la_{t_m}=0$.

By Eq. (\ref{6.4.15}), the whole hierarchy (\ref{6.4.19}) has the
standard Lax representation \beq L_{t_m}&=&[W_m,L], \label{LWMKDV}\\
W_m &=&\sum_{j=0}^{m-1} V(G_j)L^{2(m-j-1)}, \eeq where $V(G_j)$ is
given by Eq. (\ref{6.4.15}) with $G=G_j={\cal L}^j\cdot a, \ j\geq 0$,
$j\in \Z $.  Therefore we obtain the following theorem.

\begin{theorem}
The positive order hierarchy (\ref{6.4.19})
(i.e. the MKdV hierarchy)
of the spectral problem (\ref{ZS0}) possesses the Lax pair
\begin{eqnarray}
\left\{\begin{array}{l}
\psi_{x}=\left(\begin{array}{cc}
-i \la & u\\
u & i\la
\end{array}\right)\psi, \\
\psi_{t_{m}}= a\left(\begin{array}{lr}
-i \la & u\\
u & i\la
\end{array}\right)\lambda^{2(m-1)}\psi+\sum_{j=1}^{m-1}V_j\lambda^{2(m-j-1)}\psi, \ \ m=0,1,2,...,
\end{array}\right.
\end{eqnarray}
where {\normalsize \beq V_j&=& \left(\begin{array}{lr}
-i\la(uG_{j,x})^{(-1)} & \frac{1}{2}i\la
G_{j,x}-\frac{1}{4}G_{j,xx}+u(uG_{j,x})^{(-1)}\\ -\frac{1}{2}i\la
G_{j,x}-\frac{1}{4}G_{j,xx}+u(uG_{j,x})^{(-1)} & i\la(uG_{j,x})^{(-1)}
\end{array}\right),\nonumber\\
      \\
G_j&=&{\cal L}^j\cdot a, \ j\geq 0,\ j\in \Z.
\eeq} \label{Th2}
\end{theorem}
In particular, the MKdV equation (\ref{6.4.21}) has the Lax pair
\begin{eqnarray}
\left\{\begin{array}{l}
\psi_{x}=\left(\begin{array}{cc}
-i\la & u\\
u & i\la
\end{array}\right)\psi, \\
\psi_{t_{2}}=a\left(\begin{array}{lr}
-i(\la^3+\frac{1}{2}\la u^2) &
\la^2u+\frac{1}{2}i\la u_x -\frac{1}{4}u_{xx}+\frac{1}{2}u^{3}\\
\la^2u-\frac{1}{2}i\la u_x -\frac{1}{4}u_{xx}+\frac{1}{2}u^{3} &
i(\la^3+\frac{1}{2}\la u^2)
\end{array}\right)\psi.
\end{array}
\right.
\end{eqnarray}

$ $

\begin{remark}
For the spectral problem (\ref{6.4.0}), if we use the usual method
\cite{ZS,TM}, i.e. the method of finite power expansion with respect
to spectral parameter $\la$, then no isospectral evolution equations
of Eq. (\ref{6.4.0}) can be obtained.

However, we here present the MKdV hierarchy (\ref{6.4.19}) purely by
 the Lenard's operators pair satisfying Eq. (\ref{6.4.2}). Due to
 containing the spectral gradient $\nabla\la$ in Eq. (\ref{6.4.2}),
 this procedure of generating evolution equations from a given
 spectral problem is called the spectral gradient method. 
\end{remark}
In this method, how to determine a pair of Lenard's operators
associated with the given spectral problem mainly depends on the
concrete forms of spectral problems and spectral gradients, and some
computational techniques. From this method, we can furthermore
derive the negative order MKdV hierarchy, which is displayed below.

\section{The negative order MKdV hierarchy and Lax representation}
\setcounter{equation}{0}

Let us now give the negative order MKdV hierarchy through considering
the kernel element of Lenard's operator $K$.  The kernel of operator
$K$ has the following three seed functions: \beq \bar{G}^1_{-1} & = &
\left(e^{-2u^{(-1)}}\right)^{(-1)}, \label{6.4.34}\\ \bar{G}^2_{-1} &
= & \left(e^{2u^{(-1)}}\right)^{(-1)},
    \label{6.4.35}\\
 \bar{G}^3_{-1} & = & \left(e^{-2u^{(-1)}}\right)^{(-1)}
\left(e^{2u^{(-1)}}\right)^{(-1)}, \label{6.4.353} \eeq whose all
possible linear combinations form the whole kernel of $K$.  Let
$\bar{G}_{-1}\in Ker \ K$, then \be
\bar{G}_{-1}=\sum_{k=1}^3a_k\bar{G}^k_{-1} \label{G-1} \ee where
$a_k=a_k(t_n), \ k=1,2,3,$ are three arbitrarily given
$C^{\infty}$-functions with respect to variables $t_n \ (n<0,n\in
\Z)$, but independent of $x$.  Therefore, $\bar{G}_{-1}$ directly
generates the isospectral ($\la_{t_m}=0$) negative order hierarchy of
nonlinear evolution equations for the spectral problem (\ref{ZS0})
\beq u_{t_m} & = & J{\cal L}^{m+1}\cdot \bar{G}_{-1}, \ \ \ m<0, \
m\in \Z,
   \label{6.4.36}
   \eeq {\bf which is called the negative order MKdV hierarchy of
Eq. (\ref{ZS0}).}  By Theorem \ref{Th1}, the hierarchy (\ref{6.4.36})
has the standard Lax representation \beq L_{t_m}& =&\left[\bar{W}_m,
L\right]\\
\bar{W}_m&=&-\sum_{j=m}^{-1}V\left(\bar{G}_j\right)L^{2\left(m-j-1\right)},
\ \ m=-1,-2,..., \label{mkdvwm} \eeq i.e.  \beq
\left\{\begin{array}{l} \psi_{x}=\left(\begin{array}{cc} -i \la & u\\
u & i\la
\end{array}\right)\psi, \\
\psi_{t_{m}}=\sum_{j=m}^{-1}\bar{V}_j\lambda^{2\left(m-j-1\right)}\psi,
\  \ m=-1,-2,...,
\end{array}\right. \label{m4.7}
\end{eqnarray}
with {\normalsize
\beq
\bar{V}_j=
\left(\begin{array}{lr}
i\la\left(u\bar{G}_{j,x}\right)^{\left(-1\right)} &
-\frac{1}{2}i\la \bar{G}_{j,x}
-{\cal L}\cdot \bar{G}_j\\
\frac{1}{2}i\la \bar{G}_{j,x}-{\cal L}\cdot \bar{G}_j
& -i\la\left(u\bar{G}_{j,x}\right)^{\left(-1\right)}
\end{array}\right), \label{m4.8}
\eeq}
where 
$$ {\cal L}\cdot \bar{G}_j=-
\frac{1}{4}\bar{G}_{j,xx}+u\left(u\bar{G}_{j,x}\right)^{\left(-1\right)}.
 $$ In Eq. (\ref{mkdvwm}), $V\left(\bar{G}_j\right)$ and $L^{-1}$ are
given by Eq. (\ref{6.4.15}) with $G=\bar{G}_j={\cal L}^{j+1}\cdot
\bar{G}_{-1}$ and by Eq.  (\ref{6.4.6}), respectively.  Thus, all
nonlinear equations in the hierarchy (\ref{6.4.36}) are integrable.

Let us now give some special reductions of Eq. (\ref{6.4.36}).
\begin{itemize}
\item In the cases of $a_2=a_3=0;\ a_1=a_3=0; \ a_1=a_2=0,$
Eq. (\ref{6.4.36}) separately has the following representative
equations
\end{itemize}
\beq u_{t_{-1}} & =&a_1e^{-2u^{(-1)}}, \label{6.4.37}\\ u_{t_{-1}}&
 =&a_2e^{2u^{(-1)}}, \label{6.4.38}\\ u_{t_{-1}}&
 =&a_3\left(e^{-2u^{(-1)}} \left(e^{2u^{(-1)}}\right)^{(-1)}+
 e^{2u^{(-1)}}\left(e^{-2u^{(-1)}}\right)^{(-1)} \right),
 \label{6.4.383} \eeq 
which can be via the transformation $u=v_x$ 
respectively changed to 
\beq 
 v_{x,t_{-1}}&=&a_1e^{-2v},\ \ \ {\rm Liouville\  equation,} \label{6.4.391v}\\ 
 v_{x,t_{-1}}&=&a_2e^{2v},\ \ \ {\rm Liouville \ equation,} \label{6.4.392v}\\ 
 v_{t_{-1}}&=&a_3\left(e^{2v}\right)^{(-1)}\left(e^{-2v}\right)^{(-1)}, \ \ \
{\rm a\ new\ integrable \ equation}.
 \label{6.4.39v}
\eeq
They are also equivalent to the following differential equations:
\beq
& &  u_{x,t_{-1}}+2u u_{t_{-1}} =0, \label{6.4.391}\\ & & u_{x,t_{-1}}-2u
 u_{t_{-1}} =0,\label{6.4.392}\\ & &
 u_{xx,t_{-1}}-u^{-1}u_xu_{x,t_{-1}}+u^{-1}u_x-4u^2u_{t_{-1}} =0.
 \label{6.4.39}
\eeq

Apparently, Eqs. (\ref{6.4.37}), (\ref{6.4.38}), and (\ref{6.4.383})
possess the following standard Lax pairs
\begin{eqnarray}
 & & \left\{\begin{array}{l} \psi_{x}=\left(\begin{array}{cc} -i\la &
u\\ u & i\la
\end{array}\right)\psi, \\
\psi_{t_{-1}}= \bar{W}^1_{-1}\cdot\psi =\frac{1}{2}ia_1e^{-2u^{(-1)}}
             \left(\begin{array}{cc} -1 & -1\\ 1 & 1
\end{array}\right)\lambda^{-1}\psi, \\
\end{array}\right. \\
 & & \left\{\begin{array}{l} \psi_{x}=\left(\begin{array}{cc} -i\la &
u\\ u & i\la
\end{array}\right)\psi, \\
\psi_{t_{-1}}= \bar{W}^2_{-1}\cdot\psi =\frac{1}{2}ia_2e^{2u^{(-1)}}
             \left(\begin{array}{lr} 1 & -1\\ 1 & -1
\end{array}\right)\lambda^{-1}\psi, \\
\end{array}\right. \\
& & \left\{\begin{array}{l} \psi_{x}=\left(\begin{array}{cc} -i\la &
u\\ u & i\la
\end{array}\right)\psi, \\
\psi_{t_{-1}}= \bar{W}^3_{-1}\cdot\psi
=2a_3\ju{0}{1}{1}{0}\la^{-2}\psi+U_{-1}\la^{-1}\psi,
\end{array}\right.
\end{eqnarray}
respectively, where \beq U_{-1} &=& 2ia_3e^{2u^{\left(-1\right)}}
\left(e^{-2u^{\left(-1\right)}}\right)^{\left(-1\right)}
\ju{1}{-1}{1}{-1}\nonumber\\ & & + 2ia_3e^{-2u^{\left(-1\right)}}
\left(e^{2u^{\left(-1\right)}}\right)^{\left(-1\right)}
\ju{-1}{-1}{1}{1}.  \eeq

Eqs. (\ref{6.4.391}) and (\ref{6.4.392})
are two Liouville equations, and easy to see that they
 can be directly
integrated as $u^2\pm u_x=f(x), \ f(x) \in C^{\infty}(\R)$, which are
two typical Riccati equations.  They can be solved by some methods in
the theory of ordinary differential equations. But,
Eq. (\ref{6.4.39v})
or (\ref{6.4.39})
is a new integrable equation.

\begin{itemize}
\item In the case of $a_1=-\frac{1}{4}, \ a_2=\frac{1}{4}$, and
$a_3=0,$ the first equation of Eq. (\ref{6.4.36}) reads
\end{itemize}
\be u_{t_{-1}}=\frac{e^{2u^{(-1)}}-e^{-2u^{(-1)}}}{4}. \label{ut-1}
\ee We make a simple transformation \be u=\frac{1}{2}iv_x \label{uvtr}
\ee then Eq. (\ref{ut-1}) is exactly changed to {\bf the well-known
sine-Gordon equation} \be v_{x,t_{-1}}=\sin v.  \label{sinev} \ee
According to Eq. (\ref{m4.7}), the sine-Gordon equation (\ref{sinev})
possesses the following Lax pair: \beq \left\{\begin{array}{l}
\psi_x=\ju{-i\la}{\frac{1}{2}iv_x}{\frac{1}{2}iv_x}{i\la}\psi,\\
\psi_{t_{-1}}=\frac{1}{4\la}\ju{i\cos v}{\sin v}{-\sin v}{-i\cos
v}\psi,
\end{array}
\right.\label{Lax1} \eeq which has a slight difference from the usual
one given in Ref. \cite{ZMNP,Gu1}.

For Eq. (\ref{ut-1}), if we make the transformation
$u=\frac{1}{2}v_x$, then it becomes \be v_{x,t_{-1}}=\sinh v
\label{sinhv}\ee which is nothing but {\bf the well-known sinh-Gordon
equation}. By Eqs. (\ref{m4.7}) and (\ref{m4.8}), the sinh-Gordon
equation (\ref{sinhv}) has the following Lax pair \beq
\left\{\begin{array}{l}
\psi_x=\ju{-i\la}{\frac{1}{2}v_x}{\frac{1}{2}v_x}{i\la}\psi,\\
\psi_{t_{-1}}=\frac{1}{4\la}i\ju{\cosh v}{-\sinh v}{\sinh v}{-\cosh
v}\psi,
\end{array}
\right.\label{Lax2} \eeq which is also slightly different from the
usual one \cite{ZMNP,Gu1}.

\begin{remark}
In Ref. \cite{ACHM}, Alber,  Camassa, Holm and Marsden took
evolution equations of auxiliary linear system polynomial
in $\la^{-1}$ and gave the zero-curvature representation
for the Dym type hierarchy and the Camassa-Holm equation. 
Here, our starting point is the spectral problem 
and the spectral gradient instead of the auxiliary linear problem, 
then via a key operator equation
we obtain the Lax pairs.
\end{remark}

\section{A new integrable Neumann-like system}

\setcounter{equation}{0}
In Ref. \cite{QMKDV}, we presented an integrable Neumann-like
system closely associated with the positive order MKdV hierarchy
(\ref{6.4.19}),  showed  that the Neumann-like system was sent by a
 gauge transformation to an integrable Bargmann system,
and obtained the parametric solutions for the positive order MKdV hierarchy. Now, we study
the negative case.

Let $\la_1,...,\la_N$ be $N$ different spectral parameters of Eq. (\ref{6.4.0}),
 $(q_j,p_j)^T$ the spectral function corresponding to $\la_j$,  and $i^2=-1$.
Then Eq. (\ref{6.4.0}) becomes
\beqq
\left\{\begin{array}{l} 
q_{j,x}=-i\la_j q_j+up_j,\\ 
p_{j,x}=uq_j+i\la_j p_j,
\end{array}\right. 
\eeqq 
i.e. Eq. (\ref{6.4.0}) yields the following form: 
\beq
\left\{\begin{array}{l} q_x=-i\La q+up,\\ p_x=uq+i\La p,
\end{array}\right. \label{4.1}
\eeq 
where $p=(p_1,...,p_N)^T, \ q=(q_1,...,q_N)^T$, and $\La=diag(\la_1,...,\la_N)$.

Assume $\left<\La q,p\right>\not=0$, then let us restrict
Eq. (\ref{4.1}) onto the following symplectic submanifold $\mathbb{M}$
in $\R^{2N}$ \beq \mathbb{M}=\left\{(q,p)\in \R^{2N}\left|\ F\equiv
\left<q,q\right>-\left<p,p\right>-\frac{1}{4}=0,\right.\right.\nonumber\\
\left.G\equiv \frac{1}{2}\left(\left<\La q,q\right>+\left<\La
p,p\right>\right) =0\right\}.\label{4.2}\eeq Then, on $\mathbb{M}$ we
obtain the following constraint \beq
u=i\frac{\left<\La^2q,q\right>-\left<\La^2p,p\right>}{2\left<\La
q,p\right>}. \label{4.3} \eeq

Thus, under Eq. (\ref{4.3}) the MKdV spectral problem (\ref{6.4.0}) is
nonlinearized as the following nonlinear system \beq
\left\{\begin{array}{l} q_x=-i\La
q+i\frac{\left<\La^2q,q\right>-\left<\La^2p,p\right>}{2\left<\La
q,p\right>} p,\\
p_x=i\frac{\left<\La^2q,q\right>-\left<\La^2p,p\right>}{2\left<\La
q,p\right>}q+i\La p,
\end{array}\right. \label{4.4}
\eeq which is called {\bf a restricted MKdV flow of the spectral
problem (\ref{6.4.0}) on $\mathbb{M}$.}
\begin{remark}
Because $dF, \ dG $ are independent everywhere on $\M$ and
their determinant $\det(\{F,G\})= \left<\La q,p\right>\not=0$, 
   $\mathbb{M}$ is therefore a symplectic submanifold
in $\R^{2N}$ \cite{Gu1}. Apparently, $\M $ is not the usual tangent bundle, i.e.
$\mathbb{M}\not=TS^{N-1}= \left\{\left.(q,p)\in \R^{2N}\right|\
\tilde{F}\equiv \left<q,q\right>-1=0,
\tilde{G}\equiv
\left<q,p\right>=0\right\}$, thus  Eq. (\ref{4.3}) does not coincides
with the usual Neumann constraint \cite{Moser, Kn} on $TS^{N-1}$.
If we strongly impose Eq. (\ref{4.1}) on the usual tangent bundle
$TS^{N-1}= \left\{\left.(q,p)\in \R^{2N}\right|\
\tilde{F}\equiv \left<q,q\right>-1=0,\right.\\ \tilde{G}\equiv \left.
\left<q,p\right>=0\right\}$, then we have no constraints except for
$u=0$ which is of course meaningless.  So, 
 Eq. (\ref{4.4}) has no link to the standard Neumann system and is
 therefore
 a new
kind of Neumann-like system.
\end{remark}

In order to prove the integrability of the restricted flow (\ref{4.4})
on $\mathbb{M}$, we introduce the Dirac bracket \be
\{f,g\}_D=\{f,g\}+\frac{1}{2\left<\La
q,p\right>}\left(\{f,F\}\{G,g\}-\{f,G\}\{F,g\}\right)
 \label{Dirac}\ee
which is easily proven to be bilinear, skew-symmetric
and satisfy the Jacobi identity.

Let us now consider a very simple Hamiltonian function \be
H=-i\left<\La q,p\right> \label{4.5} \ee together with independent
functions {\normalsize \beq F_m&=&\frac{1}{8}\left( \left<\La^{2m}
p,p\right>-\left<\La^{2m} q,q\right>\right)\nonumber\\ &
&+\frac{1}{4}\sum_{j=m}^{-2} \left(\left<\La^{2(j+1)}
q,q\right>-\left<\La^{2(j+1)} p,p\right>\right)
\left(\left<\La^{2(m-j-1)}
p,p\right>-\left<\La^{2(m-j-1)}q,q\right>\right) \nonumber\\ &
&+\frac{1}{4}\sum_{j=m}^{-1} \left|\begin{array}{cc} \left<\La^{2j+1}
q,q\right>+\left<\La^{2j+1} p,p\right> & 2\left<\La^{2(m-j)-1}
p,q\right>\\ 2\left<\La^{2j+1} q,p\right> & \left<\La^{2(m-j)-1}
q,q\right>+\left<\La^{2(m-j)-1} p,p\right>
   \end{array}\right|,\nonumber\\
    & &  m=-1,-2,.... \nonumber\\
                     \label{4.6}
\eeq}
\begin{lemma}
The inner product $\left<\frac{\pa F_m}{\pa q},\frac{\pa F_n}{\pa
p}\right>$ is symmetric with respect to $m,\ n$ ($m, n<\\ 0,m, \ n\in
\Z$). \label{l1}
\end{lemma}
{\bf Proof}: \ \ Making the derivatives of $F_m$ with respect to $q,\
p$ and directly substituting them into $\left<\frac{\pa F_m}{\pa
q},\frac{\pa F_n}{\pa p}\right>$, we have a lengthy calculation and
then know that this inner product is sum of some symmetric terms with
respect to $m,\ n$ $(m,n=-1,-2,...)$.

\begin{proposition}
\beq
\{F_m,F_n\}=\{H,F_m\}=0, \ \ m,\ n=-1,-2,....
\eeq
\end{proposition}
{\bf Proof}: \ \ Lemma \ref{l1} directly yields $\{F_m,F_n\}=0$.  As
for the second equality, a straightforward computation completes its
proof.

$ $

Through some guesswork, we find that the restricted flow (\ref{4.4})
can be expressed as the canonical Hamiltonian form
in the following theorem.
\begin{theorem}
Under the Dirac-Poisson bracket (\ref{Dirac}), the restricted MKdV
flow (\ref{4.4}) coincides with: \beq (H)_D: \ \
\left\{\begin{array}{l} q_x=\{q,H\}_D,\\ p_x=\{p,H\}_D,
\end{array}\right. \label{H_D}
\eeq
where $H$ is defined by Eq. (\ref{4.5}).
\end{theorem}

A furthermore direct calculation leads to the following lemma.
\begin{lemma}
\beq
\{F_m,F_n\}_D=\{H,F_m\}_D=0, \ \ m,\ n=-1,-2,....
\eeq
\end{lemma}
Because $F_m$ are independent, we obtain the following theorem.
\begin{theorem}
The restricted MKdV flow (\ref{4.4}) on the symplectic
submanifold $\M$ is completely integrable. Moreover,
all restricted flows $(F_m)_D$ on $\M$
\beq
(F_m)_D: \ \ \left\{\begin{array}{l}
q_{t_m}=\{q,F_m\}_D,\\
p_{t_m}=\{p,F_m\}_D,
\end{array}\right. \label{F_D}
\eeq
are integrable.
\end{theorem}

\section{Parametric solution of the negative order\\ MKdV hierarchy}

\setcounter{equation}{0}

In the following, we will consider the relation between constraint
and nonlinear equations in the negative order MKdV hierarchy
(\ref{6.4.36}). Let us start from the following setting \beq
\bar{G}_{-1} & = & \sum_{j=1}^N\la^{-2}_j\nabla\la_j,
  \label{4.10}
\eeq where $\bar{G}_{-1}$ are defined by Eq.  (\ref{G-1}), and $
\nabla\la_j=p^2_j-q^2_j$ is the functional gradient of the spectral
problem (\ref{6.4.0}) corresponding to $\la_j$.

  Let the recursion operator ${\cal L}$ act on both sides of Eq.
  (\ref{4.10}). Then, through making a choice of $J^{-1}\cdot
  0=\pa^{-1}\cdot0= \frac{1}{4}$, we obtain \beq
  \left<p,p\right>-\left<q,q\right>=\frac{1}{4}. \label{4.12} \eeq
 Doing the derivative on both sides of Eq. (\ref{4.12}) with respect to $x$ and substituting
  Eq. (\ref{4.1}) yield \beq \left<\La q,q\right>+\left<\La
  p,p\right>=0, \label{4.13} \eeq which together with Eq. (\ref{4.12})
  forms the symplectic submanifold $\M$ we need. Apparently,
  derivative for Eq. (\ref{4.13}) with respect to $x$ leads to the constraint relation
  (\ref{4.3}). 

Since the restricted Hamiltonian flows $(H)_D$ and $(F_m)_D$ are
completely integrable and their Poisson brackets $\{H,F_m\}_D=0$
($m=-1,-2,...$), their phase flows $g^x_H,\ g^{t_m}_{F_m}$ commute
\cite{AV}. Thus, we can define their compatible solution as follows:
\beq \left(\begin{array}{l} q(x,t_m)\\ p(x,t_m)
\end{array}
\right)=g^x_H g^{t_m}_{F_m}\left(\begin{array}{l}
q(x^0,t_m^0)\\
p(x^0,t_m^0)
\end{array}
\right), \ \ m=-1,-2,..., \eeq where $x^0, \ t_m^0$ are the initial
values of phase flows $g^x_H,\ g^{t_m}_{F_m}$.
\begin{theorem}
Let $q(x,t_m), \ p(x,t_m)$ be a solution of the compatible Hamiltonian
systems $(H)_D$ and $(F_m)_D$ on $\M$. Then \beq
u=i\frac{\left<\La^2q(x,t_m),q(x,t_m)\right>-\left<\La^2p(x,t_m),p(x,t_m)\right>}
{2\left<\La q(x,t_m),p(x,t_m)\right>} \label{ur} \eeq 
satisfies the
negative order MKdV hierarchy  \beq u_{t_m} & = & J{\cal L}^{m+1}\cdot
\bar{G}_{-1}, \ \ m=-1,-2,....
   \label{umkdv}
   \eeq  \label{th4}
\end{theorem}
{\bf Proof}: On one hand, the recursion operator ${\cal L}$ acts on
Eq. (\ref{4.10}) and results in the following \beq J{\cal
L}^{m+1}\cdot \bar{G}_{-1}&=&J\cdot
\left(\left<\La^{2m}p,p\right>-\left<\La^{2m}q,q\right>\right)\nonumber\\
&=&2\left(\left<\La^{2m}p,p_x\right>-\left<\La^{2m}q,q_x\right>\right)\nonumber\\
&=&2i\left(\left<\La^{2m+1}p,p\right>+\left<\La^{2m+1}q,q\right>\right).
\eeq In this procedure, Eqs. (\ref{6.4.2}) and (\ref{4.4}) are used.

On the other hand, we directly make the derivative of Eq. (\ref{ur})
with respect to $t_m$. Then we obtain \beq u_{t_m}=\frac{i}{2\left<\La
q,p\right>^2}\left( 2\left(\left<\La^2
q,q_{t_m}\right>-\left<\La^2p,p_{t_m}\right>\right) \left<\La
q,p\right> \right.\nonumber\\ \left.-\left(\left<\La^2
q,q\right>-\left<\La^2p,p\right>\right) \left(\left<\La
q,p_{t_m}\right>+\left<\La p,q_{t_m}\right>\right)\right)
 \label{ur1}
\eeq where $q=q(x,t_m),\ p=p(x,t_m)$. But, \beq q_{t_m}=\frac{\pa
F_m}{\pa p}, \ \ p_{t_m}=-\frac{\pa F_m}{\pa q}, \eeq therefore after
substituting them into Eq. (\ref{ur1}) and calculating it, we have
\beq
u_{t_m}=2i\left(\left<\La^{2m+1}p,p\right>+\left<\La^{2m+1}q,q\right>\right)
\eeq which completes the proof.

$ $

In the special case of $m=-1$, we have the following corollary.
\begin{corollary}
Let $q(x,t_{-1}), \ p(x,t_{-1})$ be a solution of the compatible
Hamiltonian systems $(H)_D$ and $(F_{-1})_D$ on $\M$. Then \beq
u=i\frac{\left<\La^2q(x,t_{-1}),q(x,t_{-1})\right>-\left<\La^2p(x,t_{-1}),p(x,t_{-1})\right>}
{2\left<\La q(x,t_{-1}),p(x,t_{-1})\right>} \label{ur2} \eeq is a
solution of the nonlinear evolution equation $
u_{t_{-1}}=\sum^3_{k=1}a_k(t_{-1})\bar{G}^k_{-1,x}$, i.e.  \beq
u_{t_{-1}}&=&a_1e^{-2u^{(-1)}} +a_2e^{2u^{(-1)}}\nonumber\\ &
&+a_3\left(e^{-2u^{(-1)}} \left(e^{2u^{(-1)}}\right)^{(-1)}+
e^{2u^{(-1)}}\left(e^{-2u^{(-1)}}\right)^{(-1)} \right), \label{utt-1}
\eeq Here $H$ is defined by Eq. (\ref{4.5}) and $F_{-1}$ is given by
\beq F_{-1}&=&\frac{1}{8}\left (
\left<\La^{-2}p,p\right>-\left<\La^{-2}q,q\right>\right)\nonumber\\ &
& -\frac{1}{4}\left<\La^{-1}(q+p),q+p\right>
\left<\La^{-1}(q-p),q-p\right>.  \eeq 
In particular, the Liouville euqations
 $v_{x,t_{-1}}=a_1e^{-2v}, \ v_{x,t_{-1}}=a_2e^{2v}$; the sine-Gordon
equation $v_{x,t_{-1}}=\sin v$, and the sinh-Gordon equation
$v_{x,t_{-1}}=\sinh v$ have the following parametric solution 
\be 
v=u^{(-1)}=i\int
\frac{\left<\La^2q(x,t_{-1}),q(x,t_{-1})\right>-\left<\La^2p(x,t_{-1}),p(x,t_{-1})\right>}
{2\left<\La q(x,t_{-1}),p(x,t_{-1})\right>}dx,
\ee
\be
v=-2iu^{(-1)}=\int
\frac{\left<\La^2q(x,t_{-1}),q(x,t_{-1})\right>-\left<\La^2p(x,t_{-1}),p(x,t_{-1})\right>}
{\left<\La q(x,t_{-1}),p(x,t_{-1})\right>}dx, \ee 
and \be
v=2u^{(-1)}=i\int
\frac{\left<\La^2q(x,t_{-1}),q(x,t_{-1})\right>-\left<\La^2p(x,t_{-1}),p(x,t_{-1})\right>}
{\left<\La q(x,t_{-1}),p(x,t_{-1})\right>}dx, \ee respectively.
\end{corollary}

By Theorem \ref{th4}, the constraint (\ref{4.3}) is a solution of the
negative order MKdV hierarchy (\ref{umkdv}).  Thus, we also {\bf call
the system $(H)_D$ (i.e. Eq. (\ref{4.4})) a negative order restricted
MKdV flow of the spectral problem (\ref{6.4.0}) on the symplectic
submanifold $\M$.  All Hamiltonian systems $(F_m)_D$
(i.e. Eq. (\ref{F_D}) derived from $(H)_D$) are therefore called the
negative order restricted flows on $\M$}.

\section{Conclusion and Comparison}

It is well-known that some traveling wave solutions or soliton
solutions for the integrable equations are available by the Inverse
Scattering Transformation. Thus, a natural question is: what is the
relationship
between the traveling wave solutions and the parametric solutions
(\ref{ur}) for the MKdV case? Because we  need  either the potential
function $u$ decaying at $\pm \infty$ or satisfying the periodic
condition in the period $T$ with respect to the variable $x$ (see the
part of Introduction) when we
do the spectral gradient calculations, we believe that both the traveling
wave solutions and the periodic or quasi-periodic solutions should be in the
formula (\ref{ur}), namely, they share a common expression (\ref{ur}).  
In a further procedure, we will consider giving an explicit expression of
Eq. (\ref{ur})
for the cases of the potential $u$ decaying at $\pm \infty$ or satisfying the periodic
condition in the period $T$.

For the positive order constrained MKdV system by the constraint $
u=\left<p,p\right>-\left<q,q\right>$, we have dealt with it in
Ref. \cite{QMKDV} where we knew this constraint is closely connected
to the positive order (i.e. usual) MKdV hierarchy (\ref{6.4.19}) and
in detail discussed the integrability of the constrained flow for the
spectral problem (\ref{6.4.0}).

A systematic approach to generate new integrable negative order
hierarchies of NLEEs can be seen in
Ref. \cite{QCS}.

\section*{Acknowledgments}

The author would like to express his sincere thanks to
Prof. Holm,  Prof. Hyman and Prof. Margolin  for their warm
invitations and enthusiastic helps.  

This research was supported by the U.S. Department of Energy (DOE) under
contracts W-7405-ENG-36 and  the Applied Mathematical Sciences Program
KC-07-01-01; and the Special Grant of National
Excellent Doctoral Dissertation of China.

                      \end{document}